\documentclass[prd,twocolumn,showpacs,amsmath,amssymb,longbibliography]{revtex4-1}
\usepackage{graphicx}
\usepackage{bm}
\usepackage{epsfig}
\usepackage{rotating}
\usepackage{subfigure}
\usepackage{booktabs}
\usepackage{mathrsfs}
\usepackage{lineno}
\usepackage{multirow}
\usepackage[utf8]{inputenc}
\usepackage[colorlinks,citecolor=blue]{hyperref}
\usepackage{url}
\usepackage{natbib}

\newcommand{\jpsi}{J/\psi}

\newcommand{\ee}{e^+e^-}
\newcommand{\MM}{\mu^+\mu^-}

\newcommand{\piz}{\pi^0}
\newcommand{\pipi}{\pi^+\pi^-}

\newcommand{\ppjpsi}{\pi^+\pi^-\jpsi}

\newcommand{\bfg}{\begin{figure}}
\newcommand{\efg}{\end{figure}}
\newcommand{\bitm}{\begin{itemize}}
\newcommand{\eitm}{\end{itemize}}
\newcommand{\bnum}{\begin{enumerate}}
\newcommand{\enum}{\end{enumerate}}
\newcommand{\btbl}{\begin{table}}
\newcommand{\etbl}{\end{table}}
\newcommand{\btbu}{\begin{tabular}}
\newcommand{\etbu}{\end{tabular}}

\newcommand{\kk}{K^+K^-}
\newcommand{\LL}{\ell^+\ell^-}
\newcommand{\beq}{\begin{equation}}
\newcommand{\edq}{\end{equation}}

\newcommand{\ipb}{pb^{-1}}

\begin{document}

\normalsize
\parskip=5pt plus 1pt minus 1pt

\title{\boldmath Search for the $\ee \to \phi\chi_{c1}(3872)$ process at BESIII}
\author{
M.~Ablikim$^{1}$, M.~N.~Achasov$^{4,c}$, P.~Adlarson$^{75}$, O.~Afedulidis$^{3}$, X.~C.~Ai$^{80}$, R.~Aliberti$^{35}$, A.~Amoroso$^{74A,74C}$, Q.~An$^{71,58,a}$, Y.~Bai$^{57}$, O.~Bakina$^{36}$
% ...
\\
\vspace{0.2cm}
(BESIII Collaboration)\\
\vspace{0.2cm} {\it
$^{1}$ Institute of High Energy Physics, Beijing 100049, People's Republic of China\\
% ...
$^{3}$ Bochum  Ruhr-University, D-44780 Bochum, Germany\\
$^{4}$ Budker Institute of Nuclear Physics SB RAS (BINP), Novosibirsk 630090, Russia\\
% ...
$^{35}$ Johannes Gutenberg University of Mainz, Johann-Joachim-Becher-Weg 45, D-55099 Mainz, Germany\\
$^{36}$ Joint Institute for Nuclear Research, 141980 Dubna, Moscow region, Russia\\
% ...
$^{57}$ Southeast University, Nanjing 211100, People's Republic of China\\
$^{58}$ State Key Laboratory of Particle Detection and Electronics, Beijing 100049, Hefei 230026, People's Republic of China\\
% ...
$^{71}$ University of Science and Technology of China, Hefei 230026, People's Republic of China\\
% ...
$^{74}$ University of Turin and INFN, (A)University of Turin, I-10125, Turin, Italy; (B)University of Eastern Piedmont, I-15121, Alessandria, Italy; (C)INFN, I-10125, Turin, Italy\\
$^{75}$ Uppsala University, Box 516, SE-75120 Uppsala, Sweden\\
% ...
$^{80}$ Zhengzhou University, Zhengzhou 450001, People's Republic of China\\
\vspace{0.2cm}
$^{a}$ Deceased\\
% ...
$^{c}$ Also at the Novosibirsk State University, Novosibirsk, 630090, Russia\\
% ...
}
}
\date{\today}
\begin{abstract}

Based on 368.5 $\rm\ipb$ of $\ee$ collision data collected at center-of-mass energies 4.914 and 4.946 GeV 
by the BESIII detector, the $\ee \to \phi\chi_{c1}(3872)$ process  
is searched for the first time. No significant signal is observed and 
the upper limits at the 90\% confidence level on the product of the Born cross section  
$\sigma(\ee \to \phi\chi_{c1}(3872))$ 
and the branching fraction $\mathcal{B}[\chi_{c1}(3872)\to\pipi\jpsi]$ at 4.914 and 4.946 GeV
are set to be 0.85 and 0.96 pb, respectively. These measurements provide useful
information for the production of the $\chi_{c1}(3872)$ at $\ee$ colliders
and deepen our understanding about the nature of this particle.

\end{abstract}


\maketitle
The quark model categorizes hadrons into two types: mesons with a quark and an antiquark, and baryons with three quarks.
Within the framework of the theory of strong interactions, quantum chromodynamics (QCD),
it also allows the existence of more complex structures, generically called exotic hadrons.
A series of exotic hadron candidates, which cannot be accommodated by the potential model~\cite{Eichten:1978tg,Eichten:1979ms,Godfrey:1985xj},
were observed experimentally in the charmonium energy region during the past decades.
They are suggested as good candidates 
of molecule, quark-gluon hybrid, or tetraquark states~\cite{Olsen:2017bmm}. 
The well-known $\chi_{c1}(3872)$ state was first observed 
by the Belle experiment in the $ B^{\pm} \to K^{\pm} \pipi J/\psi $ decay~\cite{Belle:2003nnu}, 
and confirmed subsequently by several other experiments~\cite{CDF:2003cab,D0:2004zmu,BaBar:2004oro,LHCb:2011zzp,CMS:2013fpt,BESIII:2013fnz}. 
Ten years after its discovery, its spin-parity quantum numbers
were finally determined to be $J^{PC}=1^{++}$ by the LHCb Collaboration~\cite{LHCb:2013kgk}.  
The mass and width are determined to be $M = 3871.65 \pm 0.06$~MeV$/{c}^2$
and $\Gamma$ = 1.19 $\pm$ 0.21 MeV using a Breit-Wigner resonance model~\cite{ParticleDataGroup:2022pth}. 

Since the discovery of the $\chi_{c1}(3872)$, 
there have been tremendous efforts to understand its inner structure. 
Experimentally, there are intensive studies on the $\chi_{c1}(3872)$ decays currently. The decays of $\chi_{c1}(3872)\to $ $\pipi\jpsi$\cite{Belle:2003nnu,BESIII:2013fnz,LHCb:2020fvo}, 
$\gamma\jpsi$\cite{BESIII:2020nbj,LHCb:2014jvf,BaBar:2006fjg,Belle:2011wdj}, $\piz\chi_{c1}$\cite{BESIII:2019esk},
$\omega\jpsi$\cite{BESIII:2019qvy,BaBar:2010wfc}, 
$D^{*0} \bar{D}^{0}$~\cite{BaBar:2007cmo,Belle:2008fma,BESIII:2020nbj}
have been well observed. 
Theoretically, $\chi_{c1}(3872)$ is interpreted as a good candidate of a meson molecule~\cite{Swanson:2006st,Guo:2017jvc} 
since its mass is quite near $D^{*0} \bar{D}^{0}$ mass threshold.
On the other hand, its quantum number is $1^{++}$.
So far, the $P$-wave excited charmonium state $\chi_{c1}(2P)$ (with $J^{PC}=1^{++}$) is still missing,
and the $\chi_{c1}(3872)$ might be a good candidate for $\chi_{c1}(2P)$ since its mass is similar to the
potential model prediction~\cite{Eichten:1978tg}. Other interpretations such as a tetraquark candidate~\cite{Maiani:2004vq,Esposito:2014rxa} is also possible.
However, there is still no solid conclusion for the nature of $\chi_{c1}(3872)$.

In complement with decay, the production of $\chi_{c1}(3872)$ offers a
new window to understand its nature. In 2014, the BESIII experiment observed the
$\ee\to\gamma\chi_{c1}(3872)$ production~\cite{BESIII:2013fnz}.
What is intriguing is that the $\chi_{c1}(3872)$ might originate from the radiative transition of an excited vector state $Y(4230)$, which for the first time brings together two 
charmoniumlike states and hints commonality for their underlying nature~\cite{Olsen:2017bmm}. Recently, the 
process $\ee \to \omega\chi_{c1}(3872)$ was observed at BESIII~\cite{BESIII:2022bse}, and the production cross section shows potential enhancement near 4.75~GeV. 
In analogy to $\gamma$ and $\omega$, the vector meson $\phi$
has the same $J^{PC}$ and isospin quantum number. Therefore, the process
of $\ee\to\phi \chi_{c1}(3872)$ is expected to naturally exist. 
By investigating the relative production ratio 
$\sigma_{\ee \to\phi\chi_{c1}(3872)}/\sigma_{\ee \to\phi\chi_{c1}}$~\cite{BESIII:2022wjl}
and also comparing to
$\sigma_{\ee \to \omega\chi_{c1}(3872)}/\sigma_{\ee \to\omega\chi_{c1}}$~\cite{BESIII:2022bse,BESIII:2024ext},
we gain a deeper understanding of the $\chi_{c1}(3872)$ production and probe
the potential $\chi_{c1}(2P)$ core component in the $\chi_{c1}(3872)$ wave function~\cite{Takizawa:2012hy}. 
In addition, the decay $B^{0}_{s} \to \phi\chi_{c1}(3872)$ has also been observed~\cite{CMS:2020eiw} and has a production rate
$\mathcal{B}[B^{0}_{s} \to \phi\chi_{c1}(3872)]\approx
\mathcal{B}[B^{0} \to K^{0}\chi_{c1}(3872)]\approx
\frac{1}{2}\mathcal{B}[B^{+} \to K^{+}\chi_{c1}(3872)]~$\cite{ParticleDataGroup:2022pth}.
Together with the $\ee\to\phi\chi_{c1}(3872)$ process, a more comprehensive understanding
of $\chi_{c1}(3872)$ production will be achieved~\cite{Maiani:2020zhr}. 

In this article, we search for the $\ee \to \phi\chi_{c1}(3872)$ process
using 368.5~pb$^{-1}$ of data~\cite{BESIII:2022ulv} collected with the BESIII
detector~\cite{BESIII:2009fln} operated at the BEPCII storage ring~\cite{Yu:2016cof}. 
The $\phi$ meson is reconstructed via $\kk$ decays, while the $\chi_{c1}(3872)$ is found using $\rho^{0}\jpsi$ decays 
with $\rho^{0} \to \pipi$ and $\jpsi\to\LL$ ($\ell=e,\mu$ with close branching fraction~\cite{ParticleDataGroup:2022pth}). 
Due to the mass threshold of the $\phi\chi_{c1}(3872)$ system, only the data at the $\ee$ center-of-mass~(c.m.) energies 
$\sqrt{s}=4.914$ and 4.946 GeV is used.

The BESIII detector~\cite{BESIII:2009fln} records symmetric $e^+e^-$ collisions 
provided by the BEPCII storage ring~\cite{Yu:2016cof}
in the center-of-mass energy range from 2.0 to 4.95~GeV,
with a peak luminosity of $1 \times 10^{33}\;\text{cm}^{-2}\text{s}^{-1}$ 
achieved at $\sqrt{s} = 3.77\;\text{GeV}$. 
BESIII has collected large data samples in this energy 
region~\cite{BESIII:2020nme,Lu:2020dqs}. The cylindrical core of the BESIII detector 
covers 93$\%$ of the full solid angle and consists of a helium-based
 multilayer drift chamber~(MDC), a plastic scintillator time-of-flight
system~(TOF), and a CsI(Tl) electromagnetic calorimeter~(EMC),
which are all enclosed in a superconducting solenoidal magnet
providing a 1.0~T magnetic field.
A muon chamber (MUC) based on resistive plate chambers
with 2 cm position resolution provides information for
muon identification. 
The acceptance of charged particles and photons is 93$\%$ over $4\pi$ solid angle. 
The charged-particle momentum resolution at $1~{\rm GeV}/c$ is $0.5\%$, and the 
${\rm d}E/{\rm d}x$ resolution is $6\%$ for electrons
from Bhabha scattering. The EMC measures photon energies with a
resolution of $2.5\%$ ($5\%$) at $1$~GeV in the barrel (end cap)
region. The time resolution in the TOF barrel region is 68~ps, while
that in the end cap region is 60~ps~\cite{Li:2017jpg,Guo:2017sjt,Cao:2020ibk}.

Monte Carlo (MC) samples, simulated using {\sc geant4}-based software, are used to 
optimize the selection criteria, determine the detection efficiency and study the potential backgrounds\cite{GEANT4:2002zbu}. 
In the BESIII software framework, {\sc kkmc}~\cite{Jadach:2000ir} is the generator used to generate charmonium states by including
initial state radiation (ISR) effects and the spread of the beam energy. 
We generate 50000 signal MC events of $\ee \to \phi\chi_{c1}(3872)$  
at each c.m. energy with a phase space (PHSP) model describing 
the uniform angular distribution of the final states.
The $\chi_{c1}(3872) \to \rho^{0}\jpsi$ 
decay is described with the PHSP model.
Final state radiation of charged particles are simulated with the {\sc photos} 
package~\cite{Golonka:2005pn}. For the possible ISR effect, 
we model the $\sqrt{s}$-dependent $\ee\to\phi\chi_{c1}(3872)$ production 
cross section with a two-body phase space
($\frac{\sqrt{[s-(M_{\phi}+M_{\chi_{c1}(3872)})^{2}][s-(M_{\phi}-M_{\chi_{c1}(3872)})^{2}]}}{2\sqrt{s}}$, 
$M_{\phi}$, $M_{\chi_{c1}(3872)}$ 
is the mass of $\phi,\chi_{c1}(3872)$~\cite{ParticleDataGroup:2022pth}). Inclusive MC samples, 
with a luminosity which are ten times larger than the data sample, 
are generated at each c.m. energy to study the possible backgrounds. 
The inclusive MC sample includes the production of open charm
processes, the ISR production of vector charmonium(like) states,
and the continuum processes incorporated in {\sc
kkmc}. All particle decays are modeled with {\sc
evtgen}~\cite{Lange:2001uf,Ping:2008zz} using branching fractions 
taken from the
Particle Data Group~\cite{ParticleDataGroup:2022pth}, when available, and 
the remaining unknown charmonium decays are modeled with {\sc lundcharm}~\cite{Chen:2000tv,Yang:2014vra}.

Charged tracks detected in the MDC are required to be within a polar angle ($\theta$) 
range of $|\rm{cos\theta}|<0.93$, where $\theta$ is defined with respect to the $z$-axis,
which is the symmetry axis of the MDC. For charged tracks, the distance of closest 
approach to the interaction point (IP) 
must be less than 10\,cm along the $z$-axis, $|V_{z}|$, and less than 1\,cm
in the transverse plane, $|V_{xy}|$. For each candidate event, the pions from $\chi_{c1}(3872)$ 
decay (kaons from $\phi$ decay) and the leptons from 
$\jpsi$ decay are kinematically well separated. Charged tracks with momenta larger than
1.0 GeV/$c$ in the lab frame are assumed to be leptons, and the others are assumed to be pions or kaons. 
The energy deposited in the EMC is used to separate electrons from muons. For both
muon candidates, the deposited energy in the EMC must be less than 0.4 GeV; 
while for both electrons, it must be larger than 0.8 GeV. To separate pions
from kaons, particle identification (PID) combines measurements of the specific ionization
energy loss in the MDC~(d$E$/d$x$) and the flight time in the TOF to form 
likelihoods $\mathcal{L}(h)~(h=p,K,\pi)$ for each hadron $h$ hypothesis.
Charged kaons (pions) are identified by comparing the likelihoods for the kaon (pion) hypotheses with $\mathcal{L}(K)>\mathcal{L}(\pi)$ ($\mathcal{L}(\pi)>\mathcal{L}(K)$).

For the candidate events 
with $K^{+}K^{-} \pipi \ell^{+}\ell^{-}$ detected,
referred to as 6-track events, the
net charge is required to be zero. A four-constraint (4C) kinematic
fit imposing energy-momentum conservation is applied on the 6-track events
to improve resolution and suppress backgrounds. The kinematic fit $\chi_{4\rm{C}}^{2}$ is required to 
be less than 150. The selection criteria are optimized 
by maximizing the figure-of-merit 
\begin{displaymath}\label{1.1}
  \rm{FOM} = \epsilon_{\rm{sig}}/({\alpha}/{2}+\sqrt{N_{\rm{bkg}}}) , \tag{1.1}
\end{displaymath}
where $\epsilon_{\rm{sig}}$ is the detection efficiency from signal MC events, 
$\alpha$ is the assumed significance value which is set to 3 and $N_{\rm{bkg}}$ is the expected number of background
events obtained from inclusive MC samples. 

\begin{figure}[htbp]
    \begin{center}
    \subfigure[]{
    \includegraphics[width=0.225\textwidth]{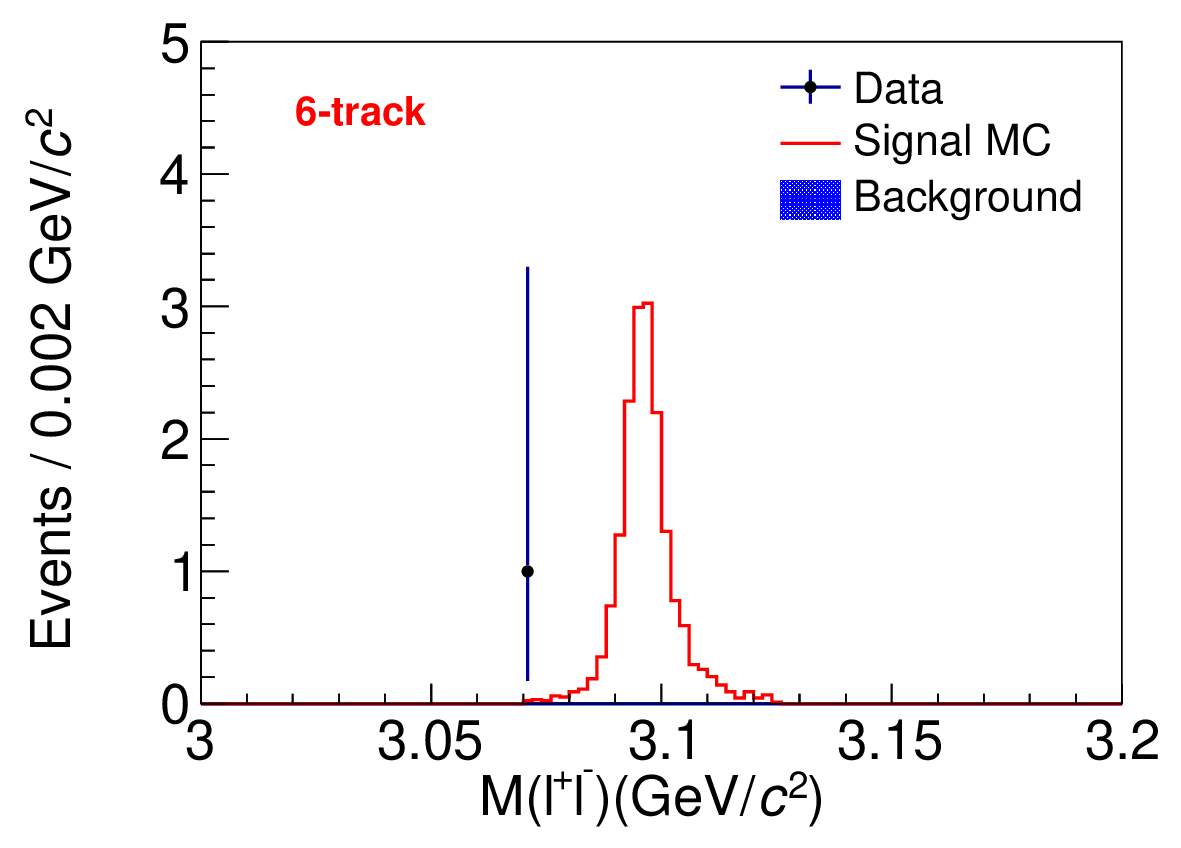}
    }
    \subfigure[]{
    \includegraphics[width=0.225\textwidth]{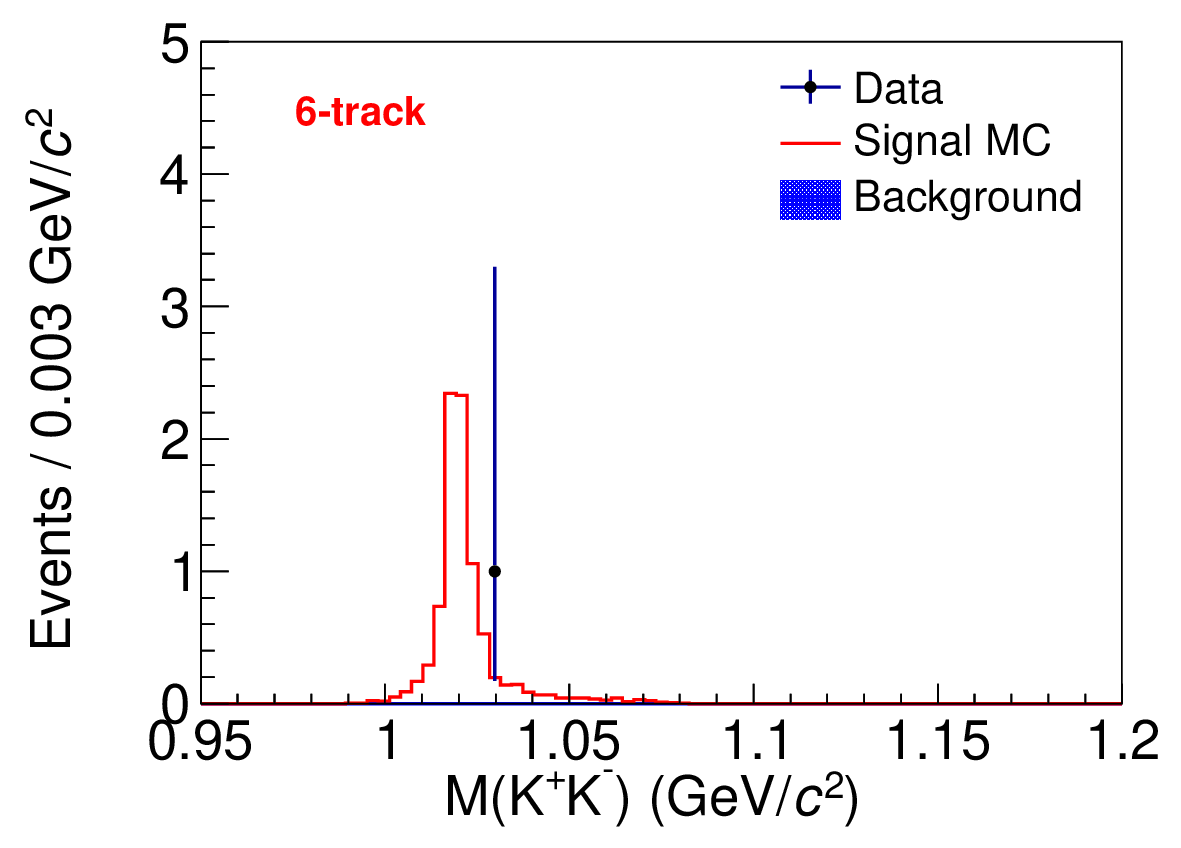}
    }
    \caption{The distributions of $M(\ell^{+}\ell^{-})$ (a) and $M(K^{+}K^{-})$ (b) for 6-track events. 
    Dots with error bars are 4.914 and 4.946 $\rm{GeV}$ data, the red histograms are 
    the signal MC sample, and the blue filled histograms are the inclusive MC sample.}
    \label{int:MM1}
    \end{center}
    \end{figure}

To select the $\jpsi$ resonance, a mass window is 
defined as $[3.070,3.125]$~GeV$/{c}^2$ (mass resolution is 6 MeV$/{c}^2$), 
which roughly covers about $\pm3\sigma$ of the $J/\psi$ signal.
The signal window of the $\phi$ resonance is set as 
$[0.980,1.080]$~GeV$/{c}^2$ (mass resolution is 7 MeV$/{c}^2$) with barely any efficiency loss according to signal MC events. 
To estimate the non-$J/\psi$ background,
the $J/\psi$ sideband regions are defined as [3.010,3.065] and $[3.130,3.185]$~GeV$/{c}^2$,
which are twice as wide as the $J/\psi$ signal region.
Due to the constraint of the mass threshold of double kaons,
the $\phi$ sideband is defined as $[1.080,1.180]$~GeV$/{c}^2$, which is
as wide as the $\phi$ signal region.
After applying all of the selection criteria, there is no 
background in the 6-track events, as indicated by the inclusive MC sample.
Figure~\ref{int:MM1} shows the invariant mass distributions of the lepton $M(\ell^{+}\ell^{-})$ and kaon $M(K^{+}K^{-})$ pairs 
for 6-track events from the full dataset.

In order to further improve the detection efficiency and thus the signal yield,
candidate events with $K^{+} \pipi \ell^{+}\ell^{-}$ 
or $K^{-} \pipi \ell^{+}\ell^{-}$ detected are also reconstructed, with one of the soft kaons 
missing due to an inefficiency.
A one-constraint (1C) kinematic fit is performed to these 5-track events,
by constraining the mass of the missing particle to the nominal mass of the 
kaon~\cite{ParticleDataGroup:2022pth}.
The kinematic fit $\chi_{1\rm{C}}^{2}$ is required to be less than 20,
which is optimized by maximizing the FOM defined in Eq. (\ref{1.1}).
Due to the absence of one track in this case, the mass resolution of the event is slightly worse 
compared to that of the 6-track events. So 
the $J/\psi$ mass window is defined as $[3.065,3.130]$~GeV$/{c}^2$ (mass resolution is 10 MeV$/{c}^2$) and
the sideband regions are [3.000,3.065] and $[3.130,3.195]$~GeV$/{c}^2$ 
for the 5-track events, which are
twice as wide as the $J/\psi$ signal region. The signal region 
of the $\phi$ resonance is set to be $ [0.980,1.080]$~GeV$/{c}^2$ (mass resolution is 8 MeV$/{c}^2$) and 
the $\phi$ sideband is defined as $[1.080,1.180]$~GeV$/{c}^2$, which is
as wide as the $\phi$ signal region.
There is a case, about $10\%$ of the total signal events, that $K^{\pm}\pi^{\mp} \pipi \ell^{+}\ell^{-}$ are detected in 
the  final state, owing to one of the soft kaon decays in the 
detector. This case is classified as 5-track events and reconstructed by missing a kaon.

\begin{figure}[htbp]
  \begin{center}
  \subfigure[]{
  \includegraphics[width=0.225\textwidth]{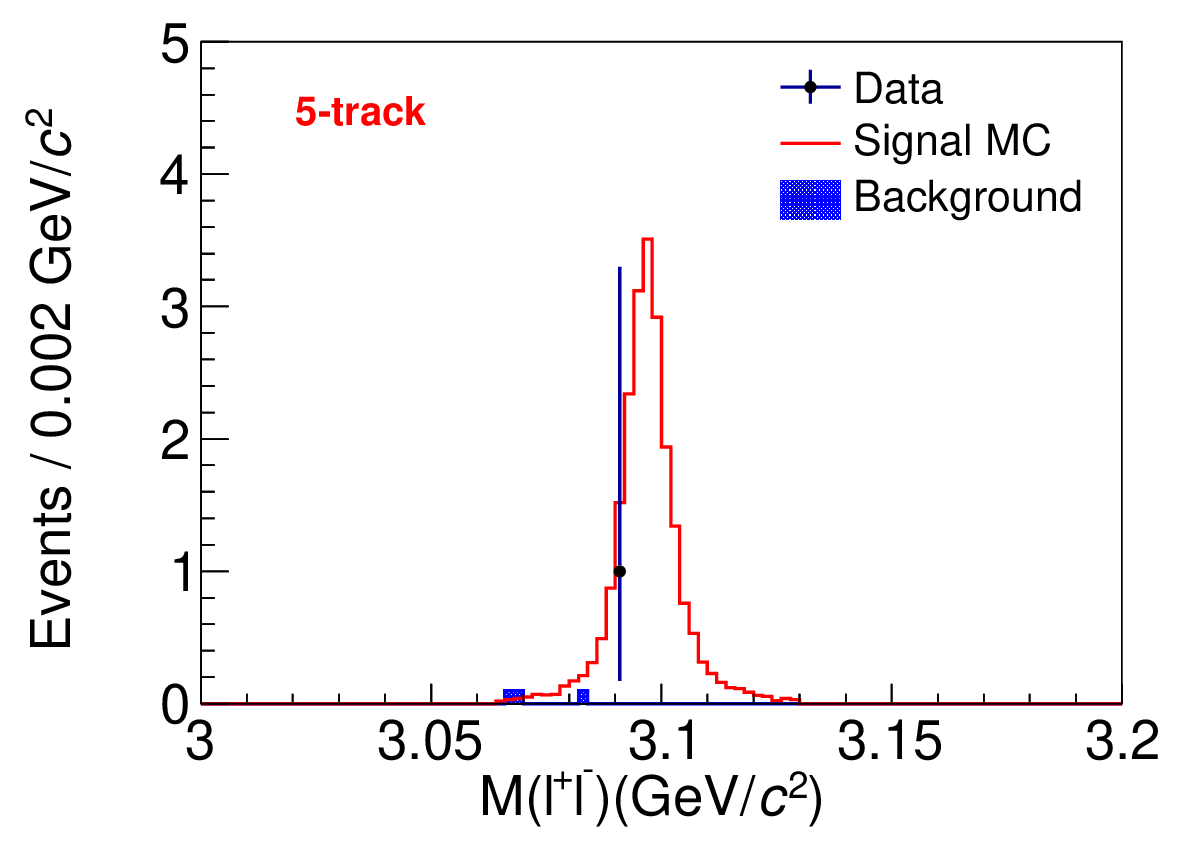}
  }
  \subfigure[]{
  \includegraphics[width=0.225\textwidth]{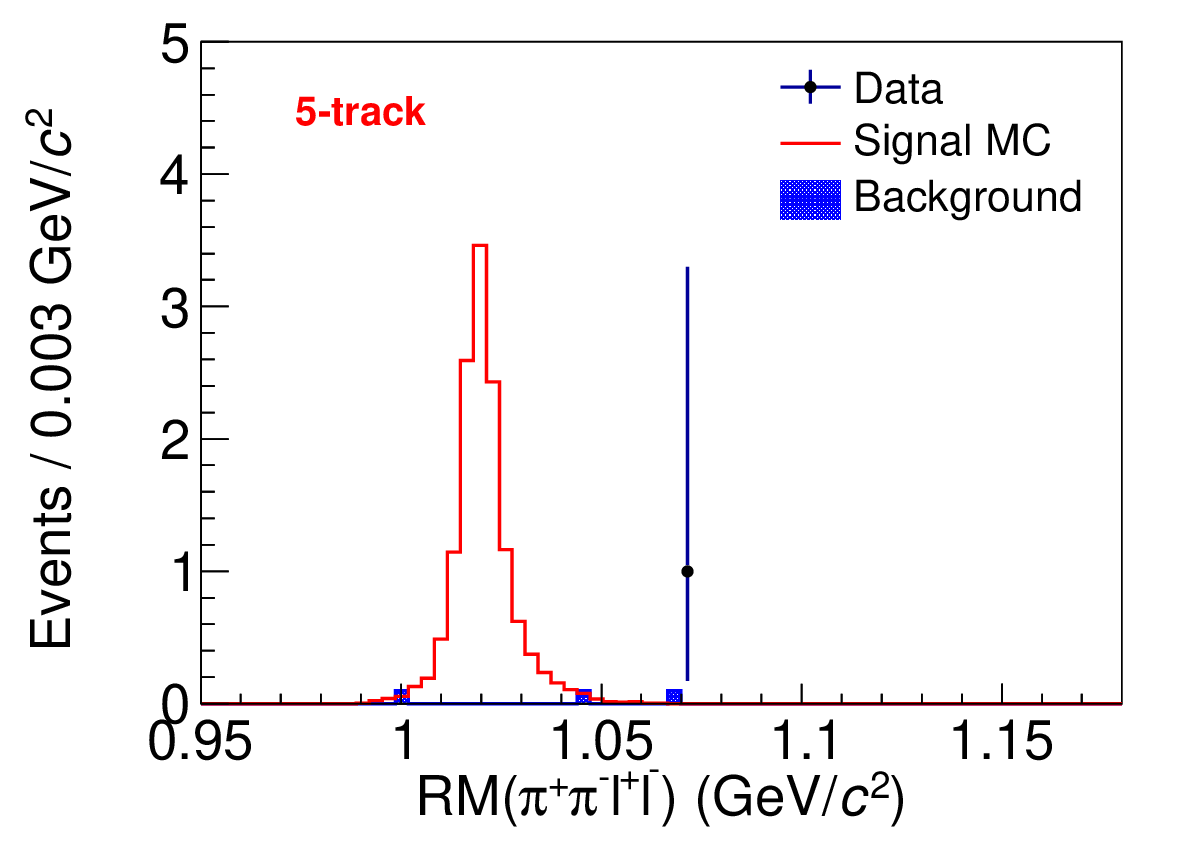}
  }
  \caption{The distributions of $M(\ell^{+}\ell^{-})$ (a) and $RM(\pipi\LL)$ (b) 
  for 5-track events. 
  Dots with error bars are 4.914 and 4.946 GeV data, the red histograms are the signal MC sample, and 
  the blue filled histograms are the inclusive MC sample.}
  \label{MM2}
  \end{center}
  \end{figure}

By analyzing the inclusve MC samples, we find there are some proton backgrounds remaining in the 5-track events.
To reduce the $p\to\mu$ misidentification background in the $\jpsi \to\MM$ channel, the MUC is used
to identify muons. At least one of the muon candidates should have a hit depth $>$ 30 cm in the MUC.
Figure~\ref{MM2} shows the $M(\ell^{+}\ell^{-})$ and $RM(\pipi\LL)$ distributions for 5-track events from the full data samples,
where $RM(\pipi\LL) = \sqrt{(P_{\ee}-P_{\pipi\LL})^{2}}$ is the recoil mass from the $\pipi\LL$ system,
$P_{\ee}$ and $P_{\pipi\LL}$ denoted the four-momenta of the initial colliding beams and the $\pipi\LL$ system.

\begin{figure}[htbp]
  \begin{center}
    \subfigure[]{
  \includegraphics[width=0.225\textwidth]{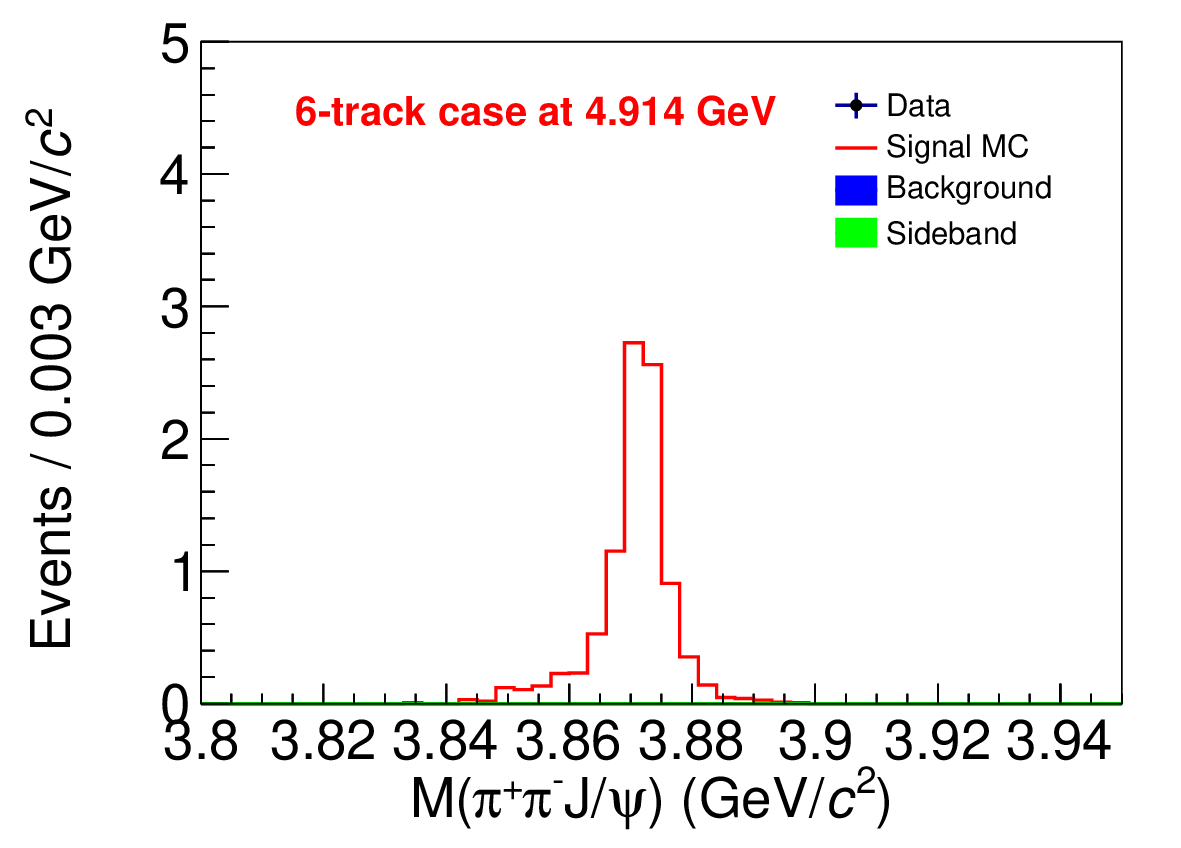}
  }
  \subfigure[]{
    \includegraphics[width=0.225\textwidth]{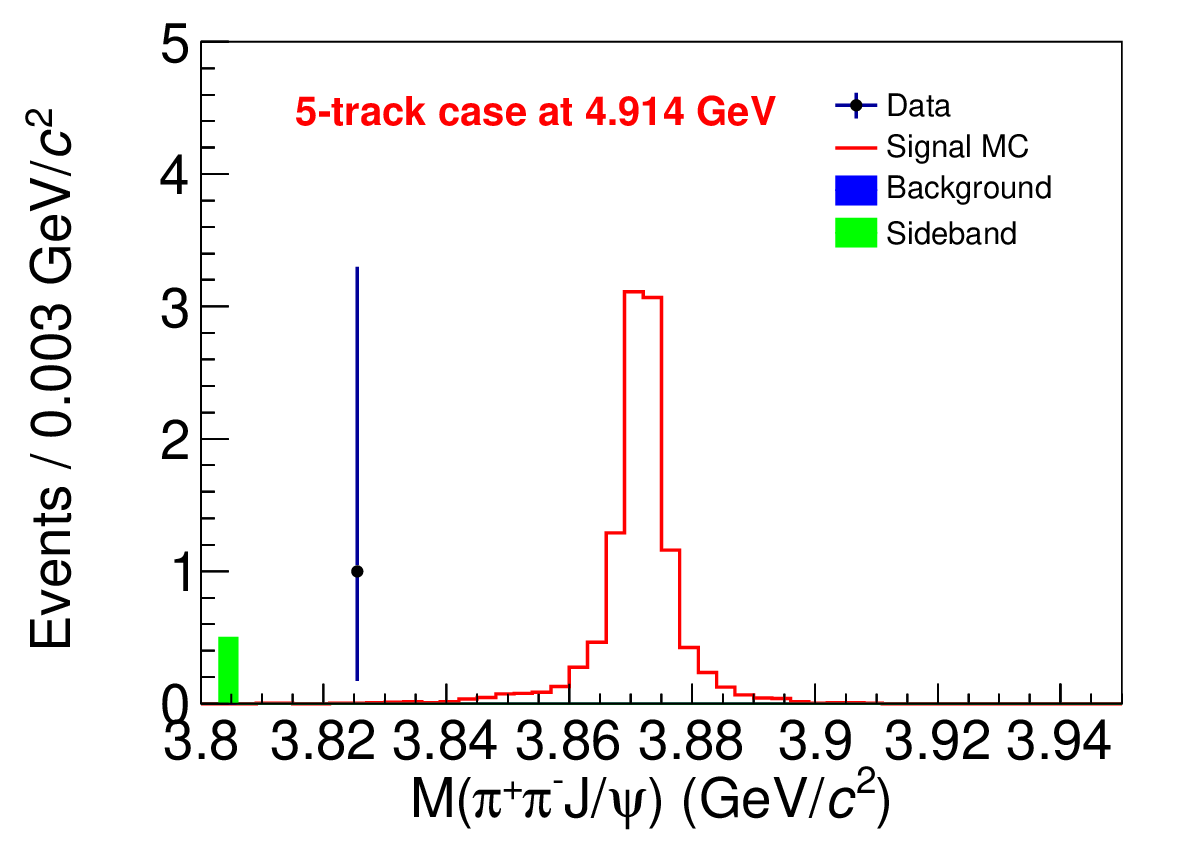}
  }
  \subfigure[]{
    \includegraphics[width=0.225\textwidth]{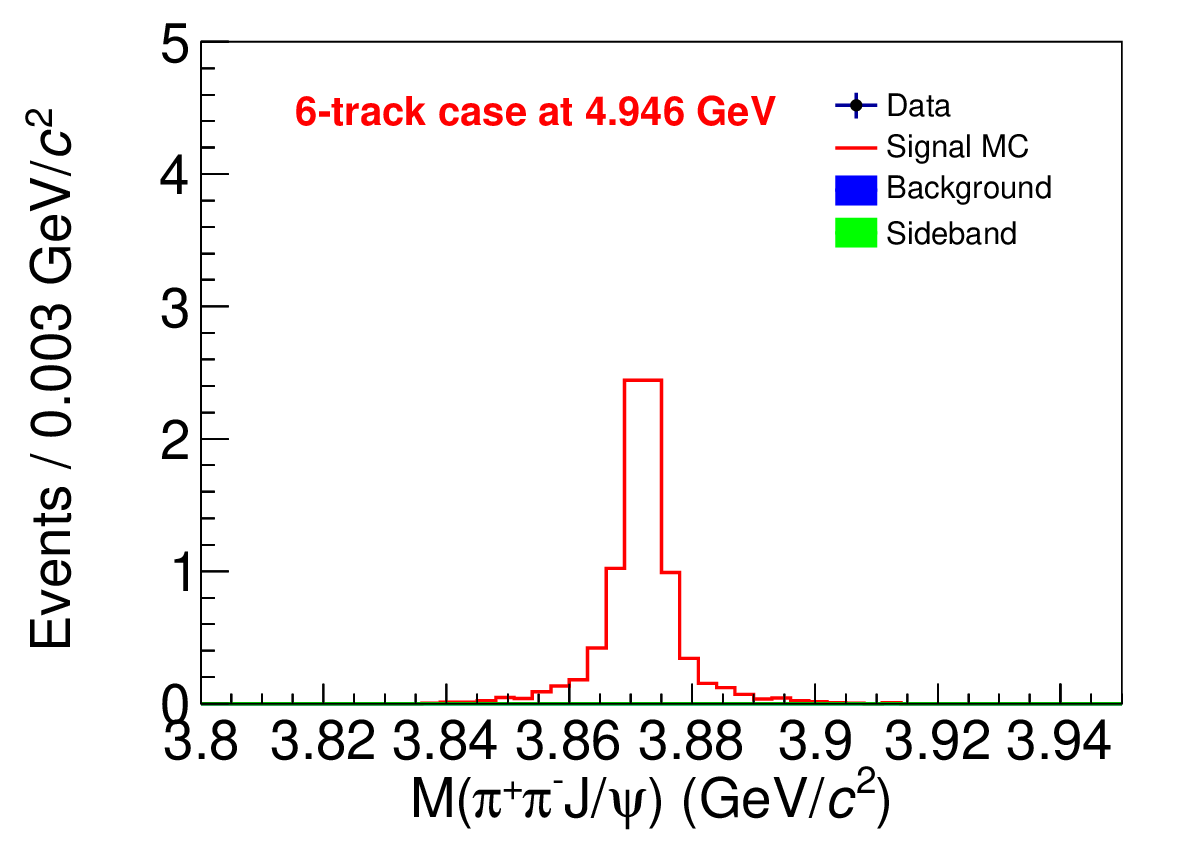}
    }
    \subfigure[]{
      \includegraphics[width=0.225\textwidth]{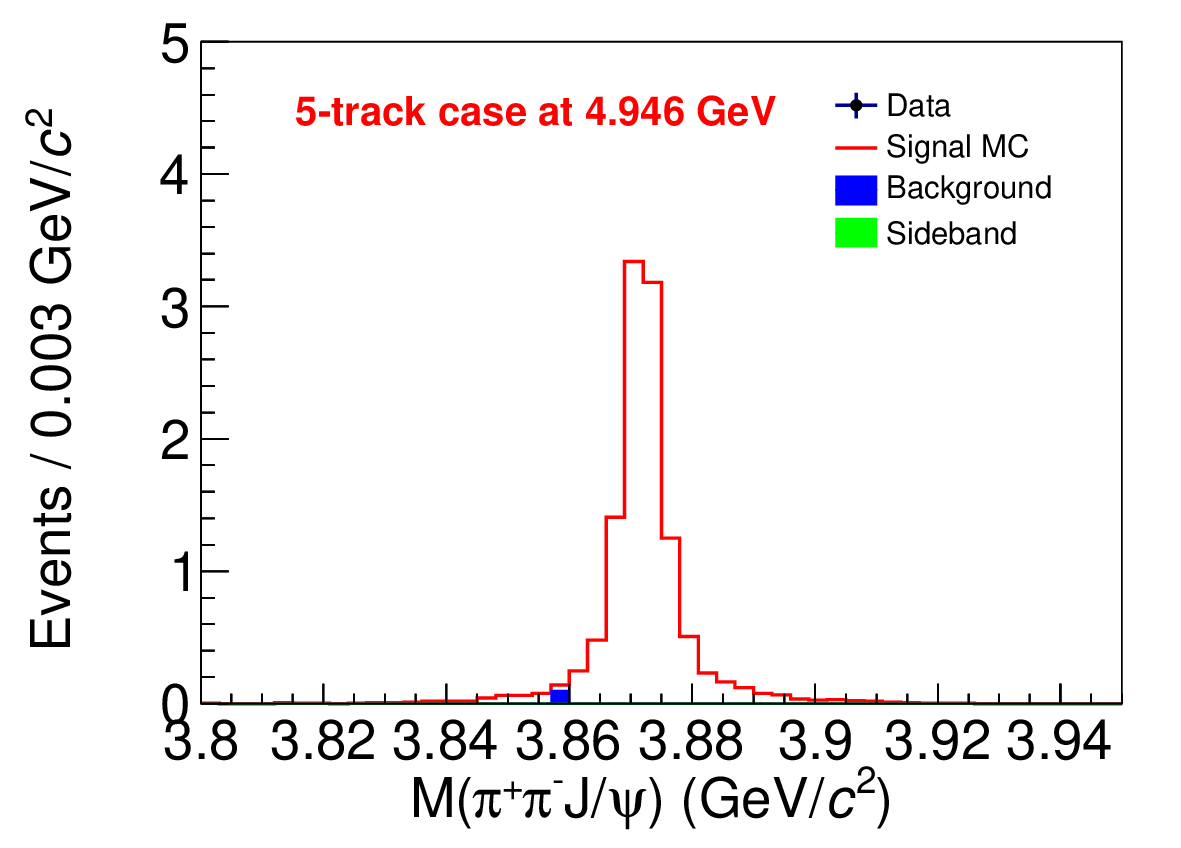}
  }
  \caption{The distributions of $M(\pipi \jpsi)$ from (a) 6-track events at $\sqrt{s} =4.914$ $\rm{GeV}$, (b) 5-track events at $\sqrt{s} =4.914\rm{GeV}$,
  (c) 6-track events at $\sqrt{s} =4.946$ $\rm{GeV}$ and (d) 5-track events at $\sqrt{s} =4.946$ $\rm{GeV}$, respectively. 
  Dots with error bars are data, the red histograms are the signal MC sample, 
  the blue filled histograms are the inclusive MC sample and 
  the green filled histograms are the $\phi-\jpsi$ 2-dimensional sideband.}
  \label{ppj}
  \end{center}
  \end{figure}

  After applying all the selection criteria, Fig.~\ref{ppj} shows 
the $M(\ppjpsi)$ distribution from 6- and 5-track events at each c.m. energy. The invariant mass $M(\pipi \jpsi)$ = $M(\pipi \ell^{+}\ell^{-})-M(\ell^{+}\ell^{-})+M(\jpsi)$ is defined, which
partly helps to cancel the resolution effect of the lepton pairs. 
Here $M(\jpsi)$ is the nominal mass of $\jpsi$ from the PDG~\cite{ParticleDataGroup:2022pth}. Through studying the inclusive MC samples at each c.m. energy, 
no dominant background survives for both 6- and 5-tracks events. 
The study of the $\jpsi$ and $\phi$ mass sideband events also shows the background level is low and so we neglect the
background.

The product of the Born cross section of $\ee \to \phi\chi_{c1}(3872)$ and $\mathcal{B}[\chi_{c1}(3872)\to\pipi\jpsi]$ at c.m. energy $\sqrt{s}$ is calculated with 
       \begin{displaymath}\label{1.2}
         \sigma(\ee \to \phi \chi_{c1}(3872))\cdot\mathcal{B}[\chi_{c1}(3872)\to\pipi\jpsi] 
        \end{displaymath}
        \begin{displaymath}
         = \frac{N_{\rm{sig}}}{\mathcal{L}_{\rm{int}}(1+\delta) \frac{1}{|1-\Pi|^{2}} \epsilon \mathcal{B}_{\rm{sub}}} , \tag{1.2}
       \end{displaymath}

 where $N_{\rm{sig}}$ is the number of signal events, $\mathcal{L}_{\rm{int}}$
is the integrated luminosity, $\epsilon$ is the selection efficiency, $\frac{1}{|1-\Pi|^{2}}=1.056$ is
the vacuum polarization factor taken from QED calculation with an accuracy of 0.05$\%$~\cite{WorkingGrouponRadiativeCorrections:2010bjp}, 
$\mathcal{B}_{\rm{sub}}$ is a product of the branching fractions $\mathcal{B}(\jpsi\to \ell^{+}\ell^{-})$ 
and $\mathcal{B}(\phi\to\kk)$, and $(1+\delta)$ is the radiative correction factor
calculated by {\sc kkmc}~\cite{Jadach:2000ir}. 

Due to no events observed in the $\chi_{c1}(3872)$ signal region, we report an upper limit on the Born cross section 
at the 90$\%$ confidence level (CL) using a frequentist method with an unbounded profile likelihood treatment~\cite{Rolke:2004mj}.
The number of signal events is determined by the $\chi_{c1}(3872)$ signal region for 6- and 5-track events, 
which is defined as $[3.86,3.88]$GeV$/{c}^2$.
The possible background estimated by the $\chi_{c1}(3872)$ sideband regions $[3.80,3.85]$ 
and $[3.90,3.95]$~GeV$/{c}^2$ from 6- and 5-track events is subtracted
(at $\sqrt{s} =4.914 \rm{GeV}$ the sideband region is defined as $[3.80,3.85]$~GeV$/{c}^2$  
due to the limitation of kinematics).
Assuming the background follows a Poisson distribution and
the efficiency (sum of 6- and 5-track events) follows a Gaussian 
distribution with a standard deviation equal to the systematic uncertainty,
the upper limit on the product of the Born cross section of $\ee \to \phi \chi_{c1}(3872)$ and
the branching fraction of $\chi_{c1}(3872)\to\pipi\jpsi$
at the 90$\%$ CL at each c.m. energy 
are measured and listed in Table~\ref{int:table6}.

\begin{table}[!h]
  \caption{The upper limit on the product of the Born cross section 
  of $\ee \to \phi \chi_{c1}(3872)$ and the branching fraction 
  of $\chi_{c1}(3872)\to\pipi\jpsi$ [denoted as $\sigma_{B}^{\rm{up}}$(pb)] at 90$\%$ CL
at each c.m. energy. 
$\sqrt{s}$~(GeV) is the c.m. energy, $\mathcal{L}_{\rm{int}}$ ($\rm\ipb$) is the integrated luminosity,
$N_{\rm{obs}}$ is the number of observed events in signal 
region, $N_{\rm{sdb}}$ is the number of observed events in the sideband region,
$N^{\rm{up}}_{\rm{signal}}$ is the upper limit on the number of observed signal at the 90$\%$ CL,
$\epsilon^{5}(\%$) and $\epsilon^{6}(\%$) are detection efficiencies 
for the 5- and 6-track events, respectively, 
$(1+\delta)$ is the radiative correction factor.}
  \centering
\begin{tabular}{cccccccccccc} \hline\hline
   $\sqrt{s}$&$\mathcal{L}_{\rm{int}}$&$N_{\rm{obs}}$&$N_{\rm{sdb}}$&$N^{\rm{up}}_{\rm{signal}}$&$(1+\delta)$&$\epsilon^{5}$&$\epsilon^{6}$&$\sigma_{B}^{\rm{up}}$\\ \hline
  4.914&208.11&0&1&1.70&0.690&19.7&2.8&0.85\\ \hline
  4.946&160.37&0&0&2.00&0.755&20.8&7.0&0.96\\ \hline\hline
   
\end{tabular}

 \label{int:table6}
\end{table}

In the cross section measurement, the systematic uncertainties are mainly 
from the luminosity measurement, branching fractions, tracking efficiency, PID efficiency, 
MUC hit depth requirement, kinematic fit,  
radiative correction factor $(1+\delta)$, MC decay model and $J/\psi$ mass window.

The uncertainty from the luminosity measurement is estimated to be less than 0.66$\%$ using
large angle Bhabha scattering events~\cite{BESIII:2022ulv}. 
The uncertainties of the decay branching fractions are quoted from the PDG~\cite{ParticleDataGroup:2022pth}. 
The uncertainty of the tracking efficiency for high momentum leptons is assigned to be 1$\%$ per track according
to the study of $\ee\to\pipi\jpsi$ at BESIII~\cite{BESIII:2016bnd}. 
In this measurement, both one-
and two kaon events are reconstructed.  
The total uncertainty of the kaon tracking efficiency
is 1.0$\%$(2.0$\%$) for 5-track(6-track) events.
Considering the uncertainty of PID efficiency is $1\%$ per kaon track at BESIII, 
the kaon PID uncertainty is 1.0$\%$(2.0$\%$) for 5-track(6-track) events, too.
The systematic uncertainty from pion tracking and PID 
efficiencies are both 1.0$\%$ per pion track~\cite{BESIII:2014bgm}.

The uncertainty of the MUC hit depth is studied using the control sample of $\ee \to \MM$~\cite{BESIII:2022wjl}. The 
difference in efficiency between data and MC simulation due to
the requirement of the $\mu$ hit depth in the MUC is taken as the systematic uncertainty.
A helix parameters correction method is used to estimate the difference between data and signal MC events
caused by the kinematic fit. The difference in efficiency with and 
without correction is taken as the systematic uncertainty. 
The systematic uncertainty of the radiative correction factor is studied 
by comparing the difference between factors obtained with the two-body phase space model and with a flat cross section line shape.
The difference in $(1+\delta)\epsilon$ is taken as the uncertainty. 
To estimate the uncertainty due
to the MC model, the angular distribution of 
$\ee \to \phi\chi_{c1}(3872)$ is modeled by a 1 $\pm$ cos$^{2}(\theta)$
distribution, where the efficiency difference with respect to PHSP is taken as the systematic uncertainty.
The control sample $\ee \to \pipi \psi(3686)$ with $\psi(3686) \to \pipi \jpsi$~\cite{BESIII:2021njb} is selected
to study the system uncertainty caused by the $\jpsi$ mass window. 
The difference in efficiency between data and signal MC events due to the mass window is taken as the
systematic uncertainty.

In the measurement, two subdata samples, i.e. the 6- and 5-track events
are reconstructed. The same source of systematic uncertainties contribute to the two
subdata samples, namely the 5- and 6-track events. These are combined
by taking the weighted average of their systematic uncertainties: Eq. (\ref{1.3}) etc.
        
  \begin{displaymath}\label{1.3}
        \Delta^{2}_{\rm{tot}}= \sum_{i=1}^{2}\omega^{2}_{i}\Delta^{2}_{i} + 2 \sum_{i < j}^{2} \rm{cov}(\it{i,j}) , \tag{1.3}
        \end{displaymath}
        \begin{displaymath}
          \rm{cov}(\it{i,j}) =  \omega_{i} \omega_{j} \rm{\Delta}_{i}\Delta_{j} , \tag{1.4}
  \end{displaymath}
  \begin{displaymath} 
         \omega_{i} = \frac{\epsilon_{i}}{\sum_{i=1}^{2} \epsilon_{i}} ,   \tag{1.5}
  \end{displaymath}
where $\Delta_{\rm{tot}}$ is the average systematic uncertainty on the cross section,
while $\omega_{i}$, $\epsilon_{i}$ and $\Delta_{i}$ are the weight, efficiency and systematic uncertainty
for the $i$th subdata sample, respectively. For the soft kaon decay events ($K^{\pm}\pi^{\mp} \pipi \ell^{+}\ell^{-}$),
its weight contributes to the 5-track subdata sample for the $\jpsi$ mass window, the MUC hit depth requirement and the kinematic fit;
otherwise its weight contributes to the 6-track subdata sample.

Assuming all these sources are independent, the total systematic uncertainty in the
cross section measurement is obtained by adding them in quadrature.
Table~\ref{un1} and Table~\ref{un2} summarize all of the 
systematic sources and their contributions at 4.914 GeV and 4.946~GeV, respectively.

\begin{table}[htbp]
  \caption{Systematic uncertainties ($\%$) in the measurement of the Born cross section at 4.914 GeV.}
  \begin{center}
  \begin{tabular}{cccc}\hline\hline
    Uncertainty&5-track&6-track&Weighted average\\ \hline
    Luminosity&\multicolumn{2}{c}{0.7}&0.7\\ 
    Tracking&5.0&6.0&5.2\\
    PID &3.0&4.0&3.2\\
    $\mathcal{B}(\jpsi\to\LL)$&\multicolumn{2}{c}{0.6}&0.6\\
    $\mathcal{B}(\phi\to\kk)$&\multicolumn{2}{c}{1.0}&1.0\\
    Radiative correction&\multicolumn{2}{c}{0.7}&0.7\\
    $\jpsi$ mass window&0.1&0.1&0.1\\
    MUC hit depth&2.1&0&1.8\\
    Kinematic fit&0.7&0.4&0.7\\ 
    MC model&5.2&5.9&5.4\\\hline
    Total&-&-&8.5\\ \hline\hline
    
  \end{tabular}
  \end{center}
  \label{un1}
  \end{table}

  \begin{table}[htbp]
    \caption{Systematic uncertainties ($\%$) in the measurement of the Born cross section at 4.946 GeV.}
   \begin{center}
   \begin{tabular}{cccc}\hline\hline
        Uncertainty&5-track&6-track&Weighted average\\ \hline
         Luminosity&\multicolumn{2}{c}{0.7}&0.7\\ 
         Tracking&5.0&6.0&5.3\\
         PID &3.0&4.0&3.3\\
         $\mathcal{B}(\jpsi\to\LL)$&\multicolumn{2}{c}{0.6}&0.6\\
         $\mathcal{B}(\phi\to\kk)$&\multicolumn{2}{c}{1.0}&1.0\\
         Radiative correction&\multicolumn{2}{c}{1.9}&1.9\\
         $\jpsi$ mass window&0.1&0.1&0.1\\
         MUC hit depth&2.9&0&2.2\\
         Kinematic fit&0.6&0.3&0.5\\ 
         MC model&2.1&14.0&6.2\\\hline
         Total&-&-&9.4\\ \hline\hline
     
 \end{tabular}
\end{center}
   \label{un2}
 \end{table}
 

In summary, with a data sample corresponding to an integrated 
luminosity of 368.5 $\rm\ipb$ collected by the BESIII detector, the process of $\ee \to \phi\chi_{c1}(3872)$
is searched for the first time. No significant signal is observed and the 
upper limits at the 90\% CL on the product of the Born cross section 
$\sigma(\ee \to \phi\chi_{c1}(3872))$ 
and the branching fraction $\mathcal{B}[\chi_{c1}(3872)\to\pipi\jpsi]$ at 4.914 and 4.946 GeV
are set to be 0.85 and 0.96 pb, respectively. Considering the 
cross section $\sigma(\ee \to \phi\chi_{c1})\sim 2.6$~pb near the production threshold~\cite{BESIII:2022wjl},
we obtain a rough estimation for the production ratio $\sigma_{\phi\chi_{c1}(3872)}/\sigma_{\phi\chi_{c1}}$$<$ 9.
It is in the same order as the relative production ratio 
$\sigma_{\omega\chi_{c1}(3872)}/\sigma_{\omega\chi_{c1}}\sim 5$~\cite{BESIII:2022bse,BESIII:2024ext}.
These measurements provide important inputs to the production of $\chi_{c1}(3872)$ at $\ee$ colliders, and help constrain the possible $\chi_{c1}(2P)$ component in the
$\chi_{c1}(3872)$ wave function.
With the upgrade of the BEPCII~\cite{BESIII:2020nme} project, more data in this energy region is expected and a more comprehensive study 
of the $\chi_{c1}(3872)$ production will be achieved, which will hopefully reveal the nature of $\chi_{c1}(3872)$ state.

\acknowledgments
The BESIII Collaboration thanks the staff of BEPCII and the IHEP computing center for their strong support. This work is supported in part by National Key R\&D Program of China under Contracts No. 2020YFA0406300, No. 2020YFA0406400, No. 2023YFA1606000; National Natural Science Foundation of China (NSFC) under Contracts No. 11635010, No. 11735014, No. 11835012, No. 11935015, No. 11935016, No. 11935018, No. 11961141012, No. 12025502, No. 12035009, No. 12035013, No. 12061131003, No. 12192260, No. 12192261, No. 12192262, No. 12192263, No. 12192264, No. 12192265, No. 12221005, No. 12225509, No. 12235017; 
the Chinese Academy of Sciences (CAS) Large-Scale Scientific Facility Program; the CAS Center for Excellence in Particle Physics (CCEPP); Joint Large-Scale Scientific Facility Funds of the NSFC and CAS under Contract No. U1832207; CAS Key Research Program of Frontier Sciences under Contracts No. QYZDJ-SSW-SLH003, No. QYZDJ-SSW-SLH040; 100 Talents Program of CAS; 
Project No. ZR2022JQ02 supported by Shandong Provincial
Natural Science Foundation; supported by the China
Postdoctoral Science Foundation under Grant
No. 2023M742100; 
The Institute of Nuclear and Particle Physics (INPAC) and Shanghai Key Laboratory for Particle Physics and Cosmology; European Union's Horizon 2020 research and innovation programme under 
Marie Sklodowska-Curie grant agreement under Contract No. 894790; German Research Foundation DFG under Contracts Nos. 455635585, Collaborative Research Center CRC 1044, FOR5327, GRK 2149; Istituto Nazionale di Fisica Nucleare, Italy; Ministry of Development of Turkey under Contract No. DPT2006K-120470; National Research Foundation of Korea under Contract No. NRF-2022R1A2C1092335; National Science and Technology fund of Mongolia; National Science Research and Innovation Fund (NSRF) via the Program Management Unit for Human Resources \& Institutional Development, Research and Innovation of Thailand under Contract No. B16F640076; Polish National Science Centre under Contract No. 2019/35/O/ST2/02907; The Swedish Research Council; U. S. Department of Energy under Contract No. DE-FG02-05ER41374.

\clearpage

\bibliography{ref}

\end{document}